\documentclass[showpacs,preprintnumbers,amsmath,amssymb]{revtex4}
\usepackage{docs}

\usepackage{graphicx}
\usepackage{dcolumn}

\usepackage{bm}

\begin{document}

\title{Plane symmetric
traversable wormholes in an anti-de Sitter background}%
\author{Jos\'e P. S. Lemos}%
\email{lemos@fisica.ist.utl.pt}
\affiliation{Centro Multidisciplinar de Astrof\'{\i}sica - CENTRA,\\
Departamento de F\'{\i}sica, Instituto Superior T\'ecnico,\\
Av. Rovisco Pais 1, 1096 Lisboa, Portugal}
\author{Francisco S. N. Lobo}%
\email{flobo@cosmo.fis.fc.ul.pt} \affiliation{Centro de Astronomia
e Astrof\'{\i}sica da Universidade de Lisboa,\\
Campo Grande, Ed. C8 1749-016 Lisboa, Portugal}

\begin{abstract}

We construct solutions of plane symmetric wormholes in the
presence of a negative cosmological constant by matching an
interior spacetime to the exterior anti-de Sitter vacuum solution.
The spatial topology of this plane symmetric wormhole can be
planar, cylindrical and toroidal. As usual the null energy
condition is necessarily violated at the throat.  At the junction
surface, the surface stresses are determined.  By expressing the
tangential surface pressure as a function of several parameters,
namely, that of the matching radius, the radial derivative of the
redshift function and of the surface energy density, the sign of
the tangential surface pressure is analyzed. We then study four
specific equations of state at the junction: zero surface energy
density, constant redshift function, domain wall equation of
state, and traceless surface stress-energy tensor. The equation
governing the behavior of the radial pressure, in terms of the
surface stresses and the extrinsic curvatures, is also displayed.
Finally, we construct a model of a plane symmetric traversable
wormhole which minimizes the usage of the exotic matter at the
throat, i.e., the null energy condition is made arbitrarily small
at the wormhole throat, while the surface stresses on the junction
surface satisfy the weak energy condition, and consequently the
null energy condition.  The construction of these wormholes does
not alter the topology of the background spacetime (i.e.,
spacetime is not multiply-connected), so that these solutions can
instead be considered domain walls. Thus, in general, these
wormhole solutions do not allow time travel.

\end{abstract}


\maketitle


\section{Introduction}

An important aspect in black hole physics is that they can be
formed through gravitational collapse of matter. This is indeed
the case for spherical collapse in an asymptotically flat
background.  For other backgrounds, such as asymptotically anti-de
Sitter spacetime with plane symmetry, it was found that
gravitational collapse of plane symmetric distributions of matter
also results in an event horizon \cite{lemos1}.  The event horizon
may have planar \cite{lemos2}, cylindrical \cite{lemos3} or
toroidal topology \cite{lemos3,zanchin}. Indeed, the collapse of
planar distributions of matter, in a background with a negative
cosmological constant, can form a planar black hole (or a black
membrane) violating somehow the hoop conjecture.  Upon
compactification of one or two coordinates one finds that
cylindrical black holes (or black strings), or toroidal black
holes can also form from the gravitational collapse of a
cylindrical or toroidal distribution of matter, respectively.  In
these solutions the mass parameter is a surface mass energy in the
planar case, a linear mass density in the cylindrical case, and a
mass in the toroidal case \cite{lemos1}.

A natural extension of these solutions would be to add exotic
matter to obtain plane symmetric traversable wormhole solutions,
with planar, cylindrical and toroidal topologies. These would add
to other non-spherically symmetric wormholes which have already
been considered by several authors. For instance, extending the
spherically symmetric Morris$\,$-Thorne wormholes \cite{Morris},
Visser \cite{Visser89} motivated by the aim of minimizing the
violation of the energy conditions and the possibility of a traveller
not encountering regions of exotic matter in a traversal through a
wormhole, constructed polyhedral wormholes and, in particular, cubic
wormholes. These contained exotic matter concentrated only at the
edges and the corners of the geometrical structure, and a traveller
could pass through the flat faces without encountering matter, exotic
or otherwise.  Gonz\'alez-D\'{\i}az generalized the static spherically
symmetric traversable wormhole solution to that of a (non-planar)
torus-like topology \cite{GDiaz}. This geometrical construction was
denoted as a ringhole. Gonz\'alez-D\'{\i}az went on to analyze the
causal structure of the solution, i.e., the presence of closed
timelike curves, and has recently studied the ringhole evolution due
to the accelerating expansion of the universe, in the presence of dark
energy \cite{GDiaz2}. Other interesting non-spherically symmetric
traversable spacetimes are the stationary solution obtained by Teo
\cite{Teo}, and an axially symmetric traversable wormhole solution
obtained by Kuhfittig \cite{Kuhf}. In this work we study plane
symmetric wormholes. We match an interior static and plane wormhole
spacetime to a vacuum solution with a negative cosmological constant,
i.e., to an anti-de Sitter spacetime. The properties of the junction
surface, such as the surface stresses are determined. For the wormhole
solutions quoted above, although the throat geometries differ from
solution to solution, all these spacetimes are asymptotically flat
with trivial topology, at infinity, whereas for the solutions we
analyze the infinity carries the same topology as the throat, meaning
that these solutions can also be considered as domain walls.
Wormholes with this same property are the spherical wormholes joining
two Friedmann-Robertson-Walker universes \cite{visserhoch}.

The plan of this paper is as follows: In Sec. II we present a
plane symmetric metric with a negative cosmological constant and
analyze the mathematics of embedding in order to obtain a wormhole
solution. In Sec. III, we present the Einstein equations for the
interior solution, and verify that the null energy condition is
necessarily violated at the wormhole throat. In Sec. IV we deduce
the exterior plane vacuum solution, through the Einstein
equations. In Sec. V, we match an interior plane spacetime to the
vacuum solution with a negative cosmological constant, and deduce
the surface stresses at the thin-shell. By expressing the
tangential surface pressure as a function of several parameters,
namely, that of the matching radius, the radial derivative of the
redshift function and of the surface energy density, the sign of
the tangential surface pressure  is analyzed. We also obtain a
general equation governing the behavior of the radial
pressure/tension across the junction in terms of the surface
stresses. In section VI, we construct a plane traversable model
that minimizes the null energy condition violation at the throat,
and in which the surface stresses on the thin shell satisfy the
energy conditions. Finally, we conclude in Sec. VII.


\section{The metric and the embedding}

\subsection{The metric}

In this work we will be concerned with the possibility of the
construction of traversable wormholes in static spacetimes and
with plane symmetry, in the presence of a negative cosmological
constant, $\Lambda<0$.  The plane symmetric metric will have the
following general form
\begin{equation}
ds^2=g_{tt}(r) \,dt^2+g_{rr}(r) \,dr^2+ \alpha^2\, r^2
\,(dx ^2+dy^2)  \,,
\end{equation}
where $\alpha$ is the inverse of the characteristic length of the
system, which here we adopt as given by the negative cosmological
constant, i.e., we put $\alpha^2=-\Lambda/3$. The ranges of $t$
and $r$ are $-\infty <t<+\infty$ and $r_0\leq r<+\infty$,
where $r_0$ is the radius of the wormhole throat.  The
range of the coordinates $x$ and $y$ determine the topology of the
plane symmetric metric. For the planar case, the topology of the
two-dimensional space, $t={\rm constant}$ and $r={\rm constant}$,
is $R^2$, with coordinate range $-\infty <x<+\infty$ and $-\infty
<y<+\infty$. For the cylindrical case the topology is $R\times
S^1$, with $-\infty <x<+\infty$ and $0 \leq \alpha\, y<2\pi$. For
the toroidal case the topology is $S^1\times S^1$ (i.e., the torus
$T\,^2$), with $0 \leq \alpha\, x<2\pi$ and $0 \leq \alpha\,
y<2\pi$ \cite{lemos3,zanchin,lemosreview,Lemos4}.

We will adopt a specific form of a static and plane symmetric
spacetime metric (with $G=c=1$), given by
\begin{equation}
ds^2=-e ^{2\Phi(r)} \,dt^2+\left(\alpha^2r^2-\frac{m(r)}{\alpha
\,r} \right)^{-1}\,dr^2 +\alpha^2r^2 \,(dx ^2+dy^2)
\label{metricwormholelambda} \,,
\end{equation}
where  $\Phi(r)$ and $m(r)$ are
arbitrary functions of the radial coordinate, $r$. $\Phi(r)$ is
called the redshift function, for it is related to the
gravitational redshift. The redshift function will be considered
finite throughout the spacetime, so that the appearance of event
horizons is avoided. $m(r)$ is called the form function, for
it determines the shape of the wormhole, which can be visualized
through embedding diagrams. We have kept the definitions of
Morris and Thorne \cite{Morris}.

\subsection{The embedding}

Embedding diagrams are a good tool to represent a wormhole and
extract some useful information for the choice of the form
function, $m(r)$ \cite{Misner}. For this plane symmetric wormhole
one can use the treatment of embedding carried out by Morris and
Thorne \cite{Morris}. Consider the interior wormhole geometry at a
fixed moment of time, and at fixed $x$ and $y$. The metric is
given by $ds^2=(\alpha^2 r^2-m(r)/\alpha r)^{-1}dr^2$. This
interior can then be embedded in a two-dimensional Euclidean
space, $ds^2=dz^2+dr^2$, where $z$ is the new extra coordinate.

Here, $z$ is only a function of $r$, $z=z(r)$,
and the condition for the embedding surface
is then
\begin{equation} \label{embeddingsurface}
\frac{dz}{dr}=\pm \left(\frac{1- \alpha^2 r^2+\frac{m(r)}{\alpha
r}}{\alpha^2 r^2-\frac{m(r)}{\alpha r}} \right)^{1/2} \,.
\end{equation}
To be a solution of a wormhole, the geometry has a minimum radius,
$r=r_0$, denoted as the throat, which defined in terms of the
shape function is given by
\begin{equation}
m(r_0)=\alpha^3\,r_0^3    \,, \label{defthroat}
\end{equation}
at which the embedded surface is vertical, i.e., $dz/dr
\rightarrow \infty$. As in a general wormhole solution, the radial
coordinate $r$ is ill-behaved near the throat, but the proper
radial distance, $l(r)=\pm \int_{r_0}^{r} (\alpha^2 r'^2
-m(r')/\alpha r')^{-1/2}\,dr'$, is required to be finite
throughout spacetime. This implies that the condition $\alpha^2
r^2 -m(r)/\alpha r \geq 0$ is imposed. Moreover, the numerator
inside the square root of Eq. (\ref{embeddingsurface}) is in
danger of going negative for sufficiently large $r$, so if one
wants a full interior embedding the particular wormhole solution
in study should take care of the problem. Now as for the exterior,
the present solution of a plane symmetric wormhole is
asymptotically anti-de Sitter. Equation (\ref{defthroat}) is still
valid for the exterior but now $m(r)$ is replaced by the total
mass $M$. One then sees that this form of embedding is only valid
for $r< r_{\rm me}$ where $r_{\rm me}$ gives the zero of the
numerator inside the square root, i.e., the maximum embedding
radius. This should not worry us. The importance of the embedding
is near the throat region where a special condition, the flare out
condition, should be obeyed. Of course, this treatment has the
drawback of being highly coordinate dependent. For a covariant
treatment see Hochberg and Visser \cite{Hochberg1,Hochberg2}.
Indeed, to be a solution of a wormhole the imposition that the
throat flares out is necessary. Mathematically, this flaring-out
condition entails that the inverse of the embedding function,
$r(z)$, must satisfy $d^2r/dz^2>0$ at or near the throat,
$m(r_0)=\alpha^3 r_0^3$. Differentiating $dr/dz=\pm
\left(\frac{\alpha^3 r^3- m(r)}{\alpha r-\alpha^3 r^3
+m(r)}\right)^{1/2}$ with respect to $z$, we have
\begin{equation}
\frac{d^2r}{dz^2}=\frac{\alpha(2\alpha^3 r^3-m'r+m)}{2
\left(\alpha r-\alpha^3r^3+m \right)^2}>0 \label{flareout}\,.
\end{equation}
Equation (\ref{flareout}) is the flaring-out condition, implying
that at the throat, $r=r_0$ or $m(r_0)=\alpha^3 r_0^3$, we have
the important condition
\begin{equation}
m'(r_0)<3\alpha^3 r_0^2   \label{flarecondition}\,,
\end{equation}
which will play a fundamental role in the analysis of the
violation of the energy conditions. Now one can draw an embedding
diagram, where the function is plotted in a $z\times r$ diagram,
see Fig. 1.

\begin{figure}[h]
\centering
  \includegraphics[width=2.4in]{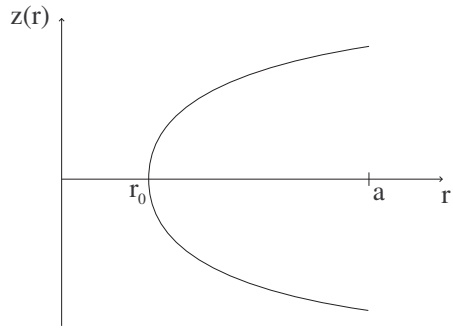}
    \caption{Embedding diagram of the plane symmetric traversable
  wormhole. The throat is situated at $r_0$, and the exotic matter
  threading the wormhole extends to a junction radius, $a$, at
  which the interior solution is matched to an exterior vacuum
  solution. It is only possible to draw the planar topology in this
embedding diagram.}
\end{figure}

Although the coordinate $x$, say, is not flat, one could think of
extending the diagram in a perpendicular direction through this
$x$ coordinate. For the planar case this extends from $-\infty$ to
$+\infty$, and thus one can see that the construction of the
wormhole does not involve a topology change. That is, the wormhole
has trivial topology.  In this sense, the wormhole solution can be
thought of as a domain wall. Compactifying the other  coordinate
$y$ does not alter the situation.  To better visualize the three
topologies we draw in Fig. 2 through pictorial diagrams, the
planar, cylindrical and toroidal topologies in the $r$ and $x$ (or
$y$) plane.
\begin{figure}[h]
\centering
  \includegraphics[width=2.1in]{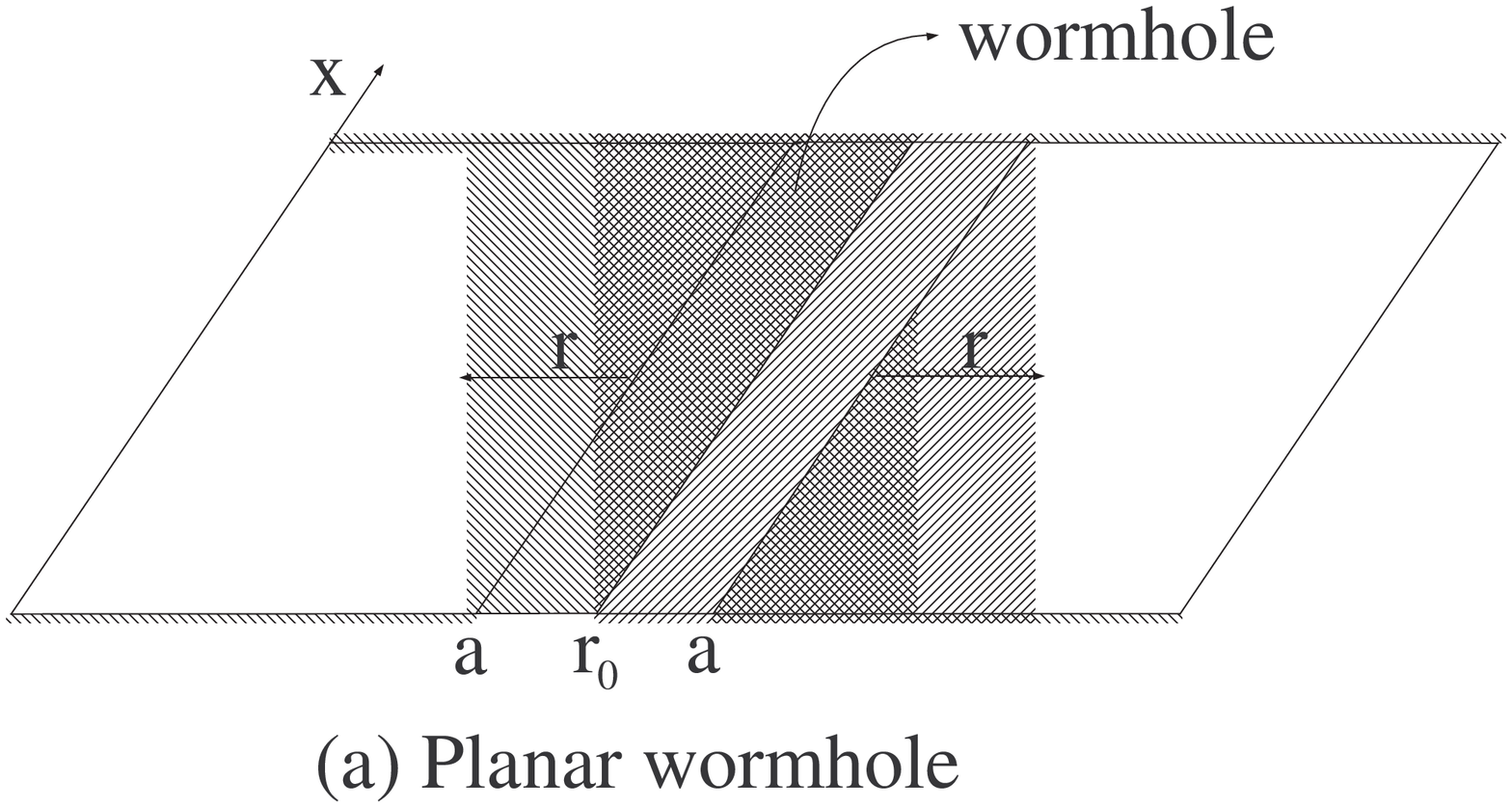}
  \hspace{0.1in}
  \includegraphics[width=1.9in]{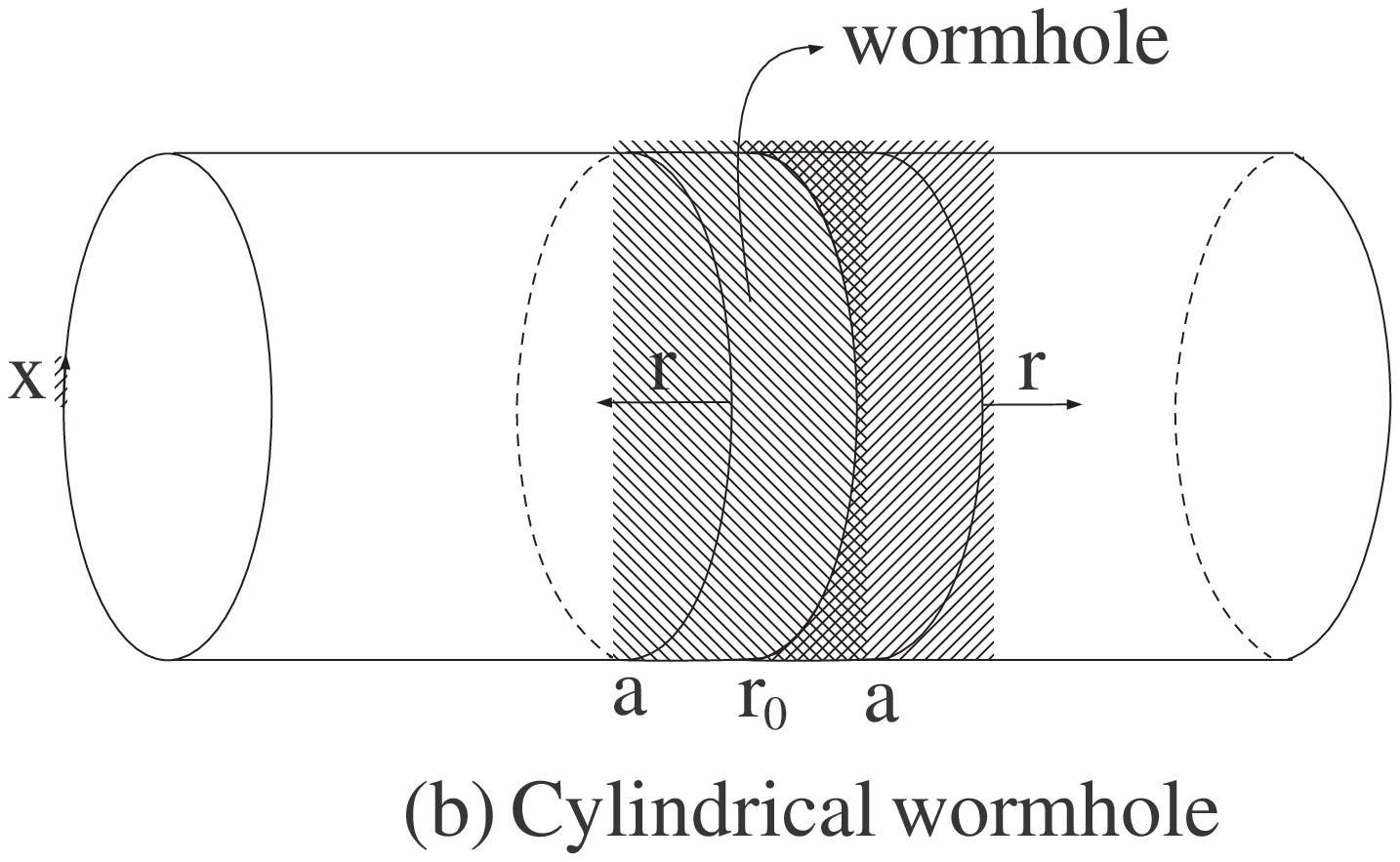}
  \hspace{0.1in}
  \centering
  \includegraphics[width=2.0in]{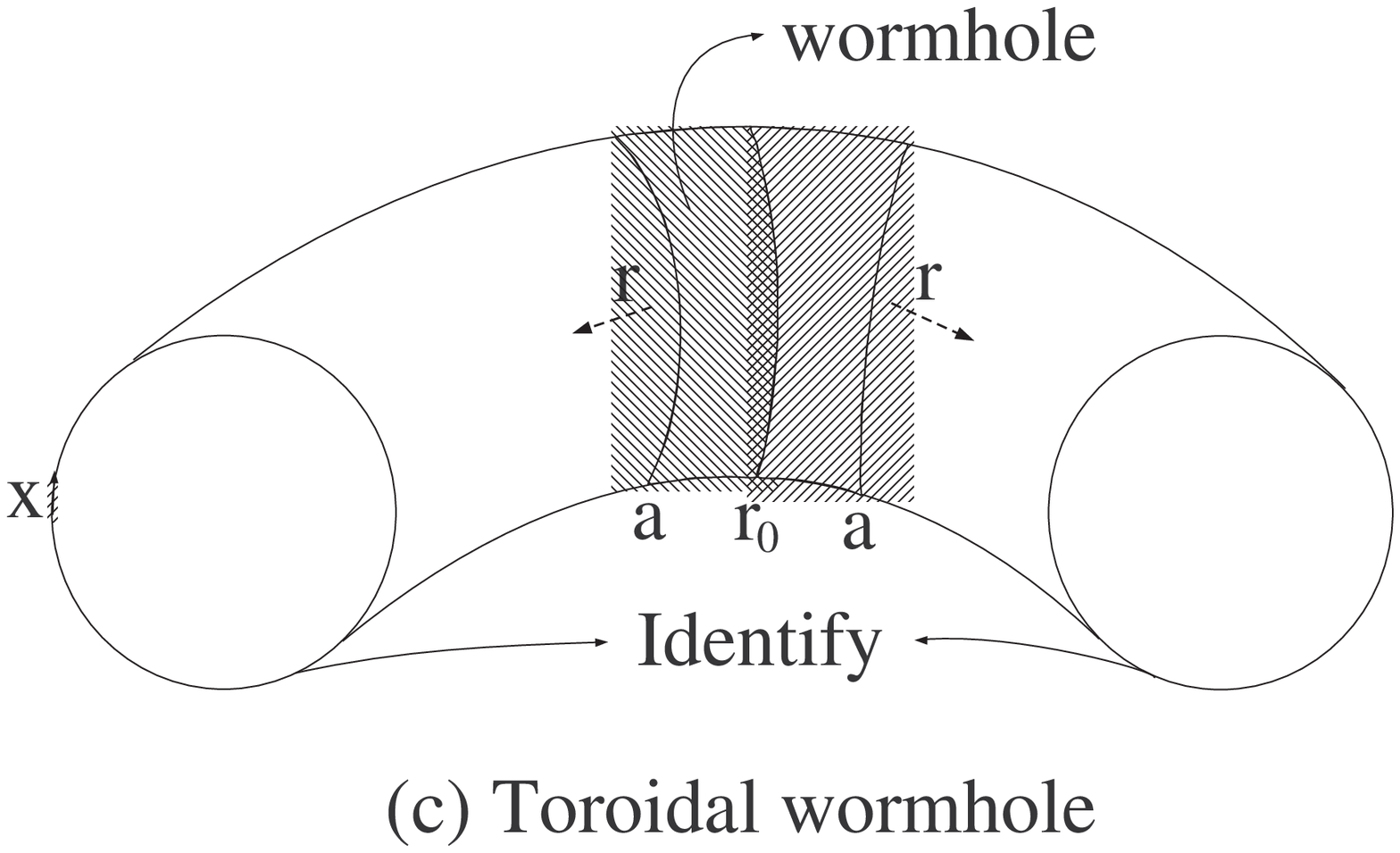}
  \caption{Pictorial diagrams of the plane symmetric traversable
  wormhole. The throat is situated at $r_0$, and the exotic matter
  threading the wormhole extends to a junction radius, $a$, at
  which the interior solution is matched to an exterior vacuum
  solution. In figure (a), we have the planar case, which can also
  be considered as a domain wall connecting two universes at the
  throat. In figure (b), one of the coordinates has been compactified,
  resulting in a cylindrical topology. In figure (c), both coordinates
  have been compactified, resulting in a planar toroidal topology.}
\end{figure}

From Fig. 2 it is clear that the construction of these wormholes
does not alter the spatial topology of the background spacetime
(i.e., spacetime is not multiply-connected), and the domain wall
interpretation can be easily recognized. This feature appears also
in the spherical wormholes joining two Friedmann-Robertson-Walker
universes \cite{visserhoch}. Note that some authors
\cite{krasnikov} do not consider as wormholes, solutions that do
not alter the topology of space.  In general, such solutions do
not allow time travel.

\section{Einstein equations with $\Lambda<0$: Interior solution}

\subsection{The equations}

Consider an orthonormal reference
frame, i.e., the proper reference frame of a set of observers who
remain at rest in the coordinate system, with $(r,x,y)$ fixed. In
the orthonormal basis the only non-zero components of the
stress-energy tensor, $T_{\hat{\mu}\hat{\nu}}$, are
$T_{\hat{t}\hat{t}}$, $T_{\hat{r}\hat{r}}$, and
$T_{\hat{x}\hat{x}}=T_{\hat{y}\hat{y}}$. Written in a diagonal
form, $T_{\hat{\mu}\hat{\nu}}={\rm diag} \left [\rho(r),p_r
(r),p_{x}(r),p_{y}(r) \right ]$, the physical interpretation of
the respective components is: $\rho(r)$ is the energy density;
$p_r(r)$ is the radial pressure; and $p_{x}(r)$ and $p_{y}(r)$ are the
tangential pressures measured in the $\partial _{x}$ and $\partial
_{y}$ directions, respectively. We define
$p(r)=p_{x}(r)=p_{y}(r)$.

In the presence of a non-vanishing cosmological constant the
Einstein field equation is given by
$G_{\hat{\mu}\hat{\nu}}+\Lambda g_{\hat{\mu}\hat{\nu}}=8\pi
T_{\hat{\mu}\hat{\nu}}$. Considering a negative cosmological
constant, we shall substitute $\Lambda=-3\alpha^2$ in the analysis
that follows. Using the metric, Eq. (\ref{metricwormholelambda}),
we obtain from the Einstein field equation
the following set of equations for
$\rho(r)$, $p_r(r)$ and $p(r)$,
\begin{eqnarray}
\rho(r)&=&\frac{1}{8\pi}\;\frac{m'}{\alpha r^2}    \label{rhoWHlambda}\,,\\
p_r(r)&=&\frac{1}{8\pi} \left [\left(\alpha^2 r^2-\frac{m}{\alpha
r}\right) \; \left(\frac{1}{r^2}+2\frac{\Phi
'}{r}\right)-3\alpha^2 \right]
     \label{radialpWHlambda}\,, \\
p(r)&=&\frac{1}{8\pi}\Bigg\{ \left(\alpha^2 r^2-\frac{m}{\alpha r}
\right)\Bigg[\Phi ''+ (\Phi')^2+ \frac{2\alpha^3
r^3-m'r+m}{2r(\alpha^3 r^3-m)}\Phi'
           \nonumber           \\
& &+\frac{2\alpha^3 r^3-m'r+m}{2r^2(\alpha^3
r^3-m)}+\frac{\Phi'}{r} \Bigg]-3\alpha^2 \Bigg\}
\label{lateralpWHlambda}\,.
\end{eqnarray}
By taking the derivative with respect to the radial coordinate,
$r$, of Eq. (\ref{radialpWHlambda}), and eliminating $m'$ and
$\Phi''$, given in Eq. (\ref{rhoWHlambda}) and Eq.
(\ref{lateralpWHlambda}), respectively, we obtain the following
equation
\begin{equation}
p_r'=-(\rho c^2+p_r)\Phi '-\frac{2}{r}(p-p_r)
\label{radialpderivative} \,,
\end{equation}
which can also be obtained using the conservation of the
stress-energy tensor,
$T^{\hat{\mu}\hat{\nu}}_{\;\;\;\;;\,\hat{\nu}}=0$, setting
$\hat{\mu}=r$. Equation (\ref{radialpderivative}) can be interpreted as
the relativistic Euler equation, or the hydrostatic equation for
equilibrium for the material threading the wormhole.

\subsection{The energy conditions for the interior}

We verify that the condition of Eq. (\ref{flarecondition}) entails
the violation of the null energy condition (NEC) at the throat and
of the weak energy condition (WEC) in a specific domain determined
below \cite{Morris, hawkingellis, Visser}. Explicitly the WEC
states $T_{\mu\nu}U^{\mu}U^{\nu}\geq0$, i.e., $\rho(r) \geq 0$ and
$\rho(r)+p_r(r) \geq 0$, where $U^{\mu}$ is a timelike vector and
$T_{\mu\nu}$ is the stress-energy tensor. Its physical
interpretation is that the local energy density is positive. In
the orthonormal frame, in the throat vicinity, we have
\begin{equation}
\rho(r_0)=T_{\hat{\mu}\hat{\nu}}U^{\hat{\mu}}U^{\hat{\nu}}=
\frac{1}{8\pi} \,\frac{m'(r_0)}{\alpha r^2_0}   \,,
         \label{WECthroat}
\end{equation}
The condition of Eq. (\ref{flarecondition}) implies $\rho(r_0) \geq
0$ for $0 \leq m'(r_0) < 3\alpha^3 r_0^2$, while $\rho(r_0)<0$ for
$m'(r_0) < 0$.

By continuity the NEC can be obtained from the WEC. The NEC states
that $T_{\mu\nu}k^{\mu}k^{\nu}\geq 0$, i.e., $\rho(r)+p_r(r) \geq
0$, where $k^{\mu}$ is a null vector. In the orthonormal frame,
$k^{\hat{\mu}}=(1,1,0,0)$, we have
\begin{equation}\label{NEC}
T_{\hat{\mu}\hat{\nu}}k^{\hat{\mu}}k^{\hat{\nu}}=\rho(r)+p_r(r)=
\frac{1}{8\pi}\,\left[\frac{m' -3\alpha^3 r^2}{\alpha r^2}+
\left(\alpha^2 r^2-\frac{m}{\alpha r}\right)
\left(\frac{1}{r^2}+2\frac{\Phi '}{r}\right) \right] \,,
\end{equation}
which at the throat, $m(r_0)=\alpha^3 r^3$, due to the finiteness
of $\Phi'$, reduces to
\begin{equation}
T_{\hat{\mu}\hat{\nu}}k^{\hat{\mu}}k^{\hat{\nu}}=
\frac{1}{8\pi}\;\frac{m'(r_0) -3\alpha^3 r_0^2}{\alpha r_0^2}  \,.
           \label{NECthroat}
\end{equation}
Equation (\ref{flarecondition}) implies the violation of the NEC
at the wormhole throat, i.e.,
$T_{\hat{\mu}\hat{\nu}}k^{\hat{\mu}}k^{\hat{\nu}}<0$. Matter that
violates the NEC is denoted as exotic matter.

\section{Einstein  equations with
$\Lambda<0$: Exterior vacuum solution}

In general, the solutions of the interior and exterior spacetimes
are given in different coordinate systems. Therefore, to
distinguish between both spacetimes we shall write out the
interior and exterior metrics in the following coordinate systems,
$(t,r,x,y)$ and $(\bar{t},\bar{r},\bar{x},\bar{y})$, respectively.
We consider the most general case in which the interior and
exterior cosmological constants, $\Lambda$ and $\bar{\Lambda}$,
are different. Therefore, the characteristic lengths are defined
as $\alpha^{-1}=\sqrt{-3/\Lambda}$ and
$\bar{\alpha}^{-1}=\sqrt{-3/\bar{\Lambda}}$, respectively. The
spacetime geometry for a vacuum exterior region is simply
determined considering a null stress-energy tensor,
$T_{\hat{\mu}\hat{\nu}}=0$, i.e.,
$\bar{\rho}(\bar{r})=\bar{p}_r(\bar{r})=\bar{p}(\bar{r})=0$. Thus,
the Einstein equations, Eqs.
(\ref{rhoWHlambda})-(\ref{lateralpWHlambda}), written in the
coordinate system, $(\bar{t},\bar{r},\bar{x},\bar{y})$, with
$\bar{m}=M$ and $e^{2\bar{\Phi}(\bar{r})}=(\bar{\alpha}^2
\bar{r}^2-M/\bar{\alpha} \bar{r})$ provide us with the exterior
vacuum solution with plane symmetry and a negative cosmological
constant, given by
\begin{equation}
\bar{ds}^2=-\left(\bar{\alpha}^2\bar{r}^2-\frac{M}{\bar{\alpha}
\bar{r}} \right)
\,d\bar{t}^2+\left(\bar{\alpha}^2\bar{r}^2-\frac{M}{\bar{\alpha}
\bar{r}}
\right)^{-1}\,d\bar{r}^2+\bar{\alpha}^2\bar{r}^2\,(d\bar{x}
^2+d\bar{y}^2)     \label{planarBH}\,.
\end{equation}
$M$ is a constant of integration and is a mass
parameter \cite{Lemos4}. Equation (\ref{planarBH}) is the metric
of a black membrane, black string, or toroidal black hole, when the
metric possesses an event horizon. Considering a positive value
for $M$, an event horizon occurs at
$\bar{r}_b=M^{1/3}/\bar{\alpha}$.
The scalar Kretschmann polynomial is given by
\begin{equation}
R^{\hat{\mu}\hat{\nu}\hat{\alpha}\hat{\beta}}
R_{\hat{\mu}\hat{\nu}\hat{\alpha}\hat{\beta}}
=24\bar{\alpha}^4+12\frac{M^2}{\bar{\alpha}^2\bar{r}^6}.
\end{equation}
showing that a singularity occurs at
$\bar{r}=0$. But this is no problem to us, since we are interested in
solutions not containing black holes.


\section{Junction conditions for plane symmetric wormholes with
an exterior $\Lambda <0$ vacuum }

\subsection{Matching of the equations I: The surface stresses}

\subsubsection{1. The surface stresses}

We shall match the interior spacetime given by
Eq. (\ref{metricwormholelambda}) to an exterior vacuum solution
with a negative cosmological constant, given by
Eq. (\ref{planarBH}), at a junction boundary $\Sigma$,
which is situated at $r=\bar{r}=a$. The intrinsic metric to
$\Sigma$ is given by
\begin{equation}
ds^2_{\Sigma}=-d\tau^2 + \alpha^2a^2 \,(dx ^2+dy^2)  \,,
\end{equation}
where $\tau$ is the proper time on $\Sigma$. Note that the
junction surface is situated outside the event horizon, i.e.,
$a>\bar{r}_b=M^{1/3}/\bar{\alpha}$, to avoid a black hole
solution.

The surface stresses at the thin shell $\Sigma$ can be determined
from the discontinuities in the extrinsic curvatures, $K_{ij}$
(see Appendix), and are given by
\begin{eqnarray}
\sigma&=&-\frac{1}{4\pi a}
\left(\sqrt{\bar{\alpha}^2a^2-\frac{M}{\bar{\alpha} a}}
-\sqrt{\alpha^2 a^2-\frac{m(a)}{\alpha a}}\;\right)  \label{surfenergy}   \,,\\
{\cal P}&=&\frac{1}{8\pi a} \left(\frac{2\bar{\alpha}^2
a^2-\frac{M}{2\bar{\alpha} a}}{\sqrt{\bar{\alpha}^2
a^2-\frac{M}{\bar{\alpha} a}}}-\zeta\,\sqrt{\alpha^2
a^2-\frac{m(a)}{\alpha a}} \,\right) \label{surfpressure}  \,,
\end{eqnarray}
where $\sigma$ is the surface energy density, and ${\cal P}$, the
tangential surface pressure on $\Sigma$. We have defined
$\zeta=1+a\Phi'(a)$ for notational convenience. In the analysis
that follows, we will only be interested in the $M>0$ case. We
shall, in general, consider a non-zero redshift function.

A particularly simple solution is given when
$(\alpha^2 a^2-m(a)/\alpha a)=(\bar{\alpha}^2
a^2-M/\bar{\alpha}a)$, in which case Eqs.
(\ref{surfenergy})-(\ref{surfpressure}) reduce to
\begin{eqnarray}
\sigma_0&=&0   \,,  \label{nullsurfenergy}  \\
{\cal P}_0&=&\frac{1}{8\pi a} \,\frac{(2-\zeta)\bar{\alpha}^2
a^2+(\zeta-\frac{1}{2})\,\frac{M}{\bar{\alpha}a}}{\sqrt{\bar{\alpha}^2
a^2-\frac{M}{\bar{\alpha}a}}}    \,.   \label{nullsurfpressure}
\end{eqnarray}
For a similar analysis in the spherically symmetric case, see
\cite{LLQ}.

\subsubsection{2. The energy conditions at the junction}

In general, the violation of the energy conditions of the surface
stresses on the boundary is verified. However, one obtains
interesting restrictions in imposing the energy conditions on the
junction surface, $\Sigma$. The WEC implies $\sigma \geq 0$ and
$\sigma + {\cal P} \geq 0$, and by continuity implies the NEC,
i.e., $\sigma + {\cal P}\geq 0$.

From Eqs. (\ref{surfenergy})-(\ref{surfpressure}), we deduce
\begin{equation} \label{sigma+P}
\sigma +{\cal P}=\frac{1}{8\pi a} \left[(2-\zeta)\,\sqrt{\alpha^2
a^2-\frac{m(a)}{\alpha a}}  + \frac{\frac{3M}{2\bar{\alpha}
a}}{\sqrt{\bar{\alpha}^2 a^2-\frac{M}{\bar{\alpha} a}}} \right]
\,.
\end{equation}
By imposing a non-negative surface energy density, $\sigma\geq 0$,
the NEC is satisfied, i.e., $\sigma +{\cal P}\geq 0$, if the
following condition is verified,
\begin{equation}  \label{energyconddomain}
\zeta \leq \frac{2\bar{\alpha}^2 a^2-\frac{M}{2\bar{\alpha}
a}}{\bar{\alpha}^2 a^2-\frac{M}{\bar{\alpha} a}} \,.
\end{equation}

\subsubsection{3. Special cases}

Taking into account Eqs. (\ref{surfenergy})-(\ref{surfpressure}),
one may express ${\cal P}$ as a function of $\sigma$, by
\begin{equation}
{\cal P}=\frac{1}{8\pi a} \,\left[\frac{(2-\zeta)\bar{\alpha}^2
a^2+(\zeta-\frac{1}{2})\,\frac{M}{\bar{\alpha}
a}}{\sqrt{\bar{\alpha}^2 a^2-\frac{M}{\bar{\alpha} a}}} -4\pi a
\zeta \sigma \right] \,.   \label{Pfunctionsigma}
\end{equation}
To analyze Eq. (\ref{Pfunctionsigma}), namely, to find domains in
which ${\cal P}$ assumes the nature of a tangential surface
pressure, ${\cal P}>0$, or a tangential surface tension, ${\cal
P}<0$, it is convenient to express Eq. (\ref{Pfunctionsigma}) in
the following compact form
\begin{equation}
{\cal P}=\frac{\bar{\alpha}}{8\pi} \,
\frac{\Gamma(\xi,\zeta,\mu)}{\sqrt{1-\xi^3}}\,, \label{compactP}
\end{equation}
with $\xi=M^{1/3}/(\bar{\alpha}a)$ and $\mu=4\pi
\sigma/\bar{\alpha}$. $\Gamma(\xi,\zeta,\mu)$ is defined as
\begin{equation}\label{Gamma}
\Gamma(\xi,\zeta,\mu)=(2-\zeta)+\left(\zeta-\frac{1}{2}\right)\xi^3-\mu
 \zeta \sqrt{1-\xi^3}    \,.
\end{equation}
One may now fix one of the parameters and analyze the sign of
$\Gamma(\xi,\zeta,\mu)$, and consequently the sign of ${\cal P}$.

The cases we shall analyze are: firstly, that of a zero surface
energy density, $\sigma=0$, i.e., $\mu=0$; secondly, a constant
redshift function, $\Phi'(r)=0$, i.e., $\zeta=1$; and finally, we
will consider two specific equations of state. The first is that
of a domain wall \cite{Visser,lake}, which is a hypersurface with
an intrinsic equation of state of the form $\sigma +{\cal P}=0$.
This equation of state is of particular interest in the present
analysis, for planar traversable wormholes may be visualized as
domain walls connecting different universes. The second equation
of state we shall consider is that of the traceless surface
stress-energy tensor, $S^{i}_{\;\,i}=0$, i.e., $-\sigma +2{\cal
P}=0$, which is an equation of state provided by the Casimir
effect with a massless field \cite{Visser}. The Casimir effect is
frequently invoked to provide exotic matter to a system violating
the energy conditions.

\bigskip
{\it (i) Null surface energy density}
\medskip

Now, considering a zero surface energy density, $\sigma=0$, i.e.,
$\mu=0$ , Eq. (\ref{Gamma}) reduces to
\begin{equation}\label{Gamma0}
\Gamma(\xi,\zeta)=(2-\zeta)+\left(\zeta-\frac{1}{2}\right)\xi^3
\,,
\end{equation}
which is analyzed in Fig. 3. Qualitatively, we verify that for
low values of $\xi$ (high $a$) and high values of $\zeta$,
$\Gamma(\xi,\zeta)$ is negative, implying a surface tension. For
low values of $\zeta$ and for all $\xi$, $\Gamma(\xi,\zeta)$ is
positive, implying a surface pressure. For high values of $\xi$
(low $a$, in the proximity of $r_b=M^{1/3}/\bar{\alpha}$) and
for all $\zeta$, a surface pressure is needed to hold the
structure against collapse.

We verify that a surface boundary, ${\cal P}=0$, i.e.,
$\Gamma(\xi,\zeta)=0$, is obtained, if
$\zeta_0=(2-\xi^3/2)/(1-\xi^3)$. This condition is equivalent to
$\Phi'(a)=(\bar{\alpha}^2a+\frac{M}{2a^2})/(\bar{\alpha}^2
a^2-\frac{M}{a})$. For a tangential surface pressure,
$\Gamma(\xi,\zeta)>0$, we have $\zeta<\zeta_0$. For a tangential
surface tension, $\Gamma(\xi,\zeta)<0$, the condition
$\zeta>\zeta_0$ is imposed. In particular, considering a constant
redshift function, i.e., $\Phi'(r)=0$, so that $\zeta=1$, we
verify that ${\cal P}$ is a tangential surface pressure, ${\cal
P}>0$.

For $\sigma=0$, we verify that the WEC is satisfied only if ${\cal
P} \geq 0$. This condition is verified in the domain $\zeta\leq
\zeta_0$, which is consistent with Eq. (\ref{energyconddomain}).

\begin{figure}[h]
  \centering
  \includegraphics[width=3.0in]{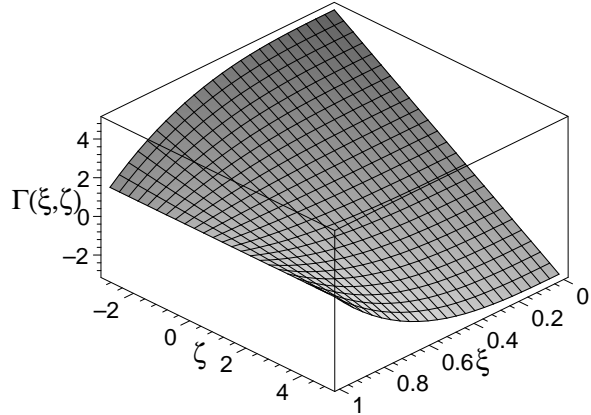}
  \caption{Considering a zero surface energy density, $\sigma=0$, i.e.,
$\mu=0$, we have defined
$\xi=M^{1/3}/(\bar{\alpha}a)$ and $\zeta=1+a\Phi'(a)$. The surface is
defined by
$\Gamma(\xi,\zeta)=(2-\zeta)+(\zeta-1/2)\,\xi^3$. One verifies that
for low values of $\xi$ (high $a$) and high values of $\zeta$,
$\Gamma(\xi,\zeta)$ is
negative, implying a surface tension, ${\cal P}<0$. Whilst for
low values of
$\zeta$ and for all $\xi$, $\Gamma(\xi,\zeta)$ is positive,
implying a surface
pressure, ${\cal P}>0$. For high vales of $\xi$ (low $a$, in
the proximity of $r_b$) and for
arbitrary values of $\zeta$, a surface
pressure is needed to hold the structure against collapse.
See text for details.}
\end{figure}

\bigskip
{\it (ii) Constant redshift function}
\medskip

Considering the specific case of a constant redshift function,
$\Phi'(r)=0$, i.e., $\zeta=1$, Eq. (\ref{Gamma}) reduces to
\begin{equation}\label{Gamma1}
\Gamma(\xi,\mu)=1+\frac{\xi^3}{2}-\mu \sqrt{1-\xi^3}    \,.
\end{equation}
This relationship is represented as the surface in Fig. $4$.
Qualitatively, considering low values of $\xi$ (high $a$) and high
positive values of $\mu$, ${\cal P}$ assumes the character of a
surface tension, i.e., ${\cal P}<0$. For negative values of the
surface energy density and for all $\xi$, ${\cal P}$ is always
positive. For extremely high values of $\xi$ (extremely low $a$,
in the proximity of $r_b=M^{1/3}/\bar{\alpha}$) and for all
$\sigma$, a surface pressure is needed to hold against collapse.

Analytically, we verify that ${\cal P}=0$ at
$\mu_0=(1+\xi^3/2)/\sqrt{1-\xi^3}$; ${\cal P}$ assumes a
tangential surface pressure, ${\cal P}>0$, for $\mu<\mu_0$, and a
tangential surface tension, ${\cal P}<0$, for $\mu>\mu_0$.

For $\zeta=1$, i.e., $\Phi'(r)=0$, by imposing $\sigma \geq 0$, we
verify that by taking into account Eq. (\ref{sigma+P}), the WEC is
immediately satisfied. If $\sigma <0$, then only the WEC is
violated, while the NEC is satisfied.

\begin{figure}[h]
  \centering
  \includegraphics[width=3.0in]{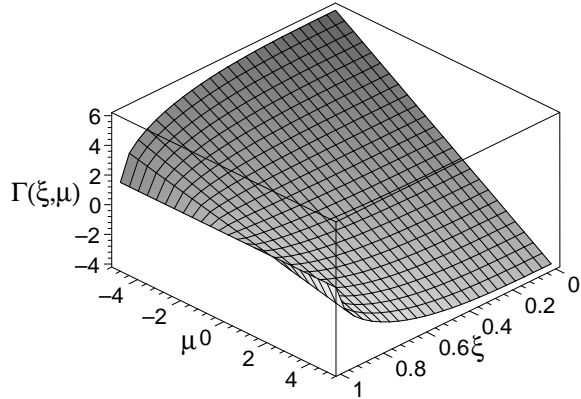}
  \caption{Considering a constant value for the redshift function,
  $\Phi'(r)=0$, i.e., $\zeta=1$, we have defined
  $\xi=M^{1/3}/(\bar{\alpha}a)$ and
  $\mu=4\pi \sigma/\bar{\alpha}$. The surface is defined by
  $\Gamma(\xi,\mu)=1+\xi^3/2-\mu \sqrt{1-\xi^3}$.
  For low values of $\xi$ (high $a$) and positive values of $\mu$,
  $\Gamma(\xi,\mu)$ is negative, implying a
  surface tension, ${\cal P}<0$. While, for negative values of
  the surface energy
  density and for all $\xi$, $\Gamma(\xi,\mu)$ is positive, implying a
  surface pressure, ${\cal P}>0$. For extremely high values of $\xi$
  (for values of $a$
  in the neighborhood of $r_b$), a surface pressure is once again
  required to hold against collapse. See text for details.}
\end{figure}

\bigskip
{\it (iii) Domain wall}
\medskip

The domain wall is a hypersurface with an intrinsic equation of
state of the form $\sigma +{\cal P}=0$ \cite{Visser,lake}. Taking
into account Eqs. (\ref{surfenergy})-(\ref{surfpressure}), from
$\sigma +{\cal P}=0$, we deduce the following condition
\begin{equation}
(\zeta-2)\sqrt{\alpha^2 a^2-\frac{m(a)}{\alpha a}} =
\frac{\frac{3M}{2 \bar{\alpha}a}}{\sqrt{\bar{\alpha}^2
a^2-\frac{M}{\bar{\alpha}a}}}   \,.
\end{equation}
The right hand side term is always positive, for by construction
we have only taken into account $M>0$. Thus, to have a solution,
this imposes the important condition $\zeta>2$.

The tangential surface pressure is given by
\begin{equation}\label{classmembraneP}
{\cal P}=\frac{1}{4\pi a}\,\frac{1}{\sqrt{\bar{\alpha}^2
a^2-\frac{M}{\bar{\alpha}a}}} \left[\bar{\alpha}^2
a^2-\left(\frac{2\zeta -1}{\zeta-2}\right)\frac{M}{2\bar{\alpha}a}
\right]    \,.
\end{equation}
Equation (\ref{classmembraneP}) may also be obtained by
substituting $\sigma=-{\cal P}$ into Eq. (\ref{Pfunctionsigma}),
and can be written in a compact form as ${\cal
P}=\frac{\bar{\alpha}}{4\pi} \,
\frac{\Gamma(\xi,\zeta)}{\sqrt{1-\xi^3}}$, with
$\Gamma(\xi,\zeta)$ given by
\begin{equation}\label{Gammaclassmem}
\Gamma(\xi,\zeta)=1-\left(\frac{2\zeta-1}{\zeta-2}\right)\frac{\xi^3}{2}
\,.
\end{equation}
$\Gamma(\xi,\zeta)$ is represented as a surface in Fig. 5.
Qualitatively, we verify that for low values of $\xi$ (high $a$)
and for all $\zeta$, $\Gamma(\xi,\zeta)$ is positive, implying a
surface pressure, ${\cal P}>0$. While, for high values of $\xi$
(low $a$) and for all $\zeta$, $\Gamma(\xi,\zeta)$ is negative,
implying a surface tension, ${\cal P}<0$. We have a null
tangential surface pressure, ${\cal P}=0$, for
$\zeta_0=(2-\xi^3/2)/(1-\xi^3)$. ${\cal P}$ assumes the character
of a surface pressure, ${\cal P}>0$, for $\zeta>\zeta_0$ and a
surface tension, ${\cal P}<0$, for $2<\zeta<\zeta_0$.

From the equation of state, $\sigma=-{\cal P}$, we verify that
$\sigma \geq 0$ if and only if ${\cal P} \leq 0$, thus satisfying
the WEC.

\begin{figure}[h]
  \centering
  \includegraphics[width=3.0in]{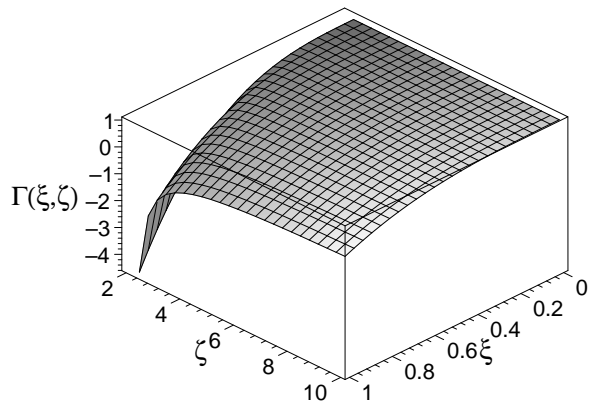}
  \caption{The surface represents the sign of ${\cal P}$ for a domain wall,
  with the equation of state $\sigma +{\cal P}=0$.
  We have defined $\xi=M^{1/3}/(\bar{\alpha}a)$ and $\zeta=1+a\Phi'(a)$.
  For low values of $\xi$ (high $a$) and for all $\zeta$,
  $\Gamma(\xi,\zeta)$ is positive, implying a surface pressure, ${\cal P}>0$.
  While, for high values of $\xi$ (low $a$) and for all $\zeta$,
  $\Gamma(\xi,\zeta)$ is negative, implying a surface tension, ${\cal P}<0$.
  See text for details.}
\end{figure}

\bigskip\bigskip\bigskip\bigskip
{\it (iv) Traceless surface stress-energy tensor}
\medskip

The traceless surface stress-energy tensor, $S^{i}_{\;\,i}=0$,
i.e., $-\sigma +2{\cal P}=0$, is a case of particular interest
\cite{Visser}. The Casimir effect with a massless field provides
one with a stress-energy tensor of this type. Usually, the Casimir
effect is theoretically invoked to provide exotic matter to the
system considered at hand. From $\sigma =2{\cal P}$ and taking
into account Eqs. (\ref{surfenergy})-(\ref{surfpressure}), we have
the following condition
\begin{equation}
(1+\zeta)\sqrt{\alpha^2 a^2-\frac{m(a)}{\alpha a}} =
\frac{3\left(\bar{\alpha}^2 a^2-\frac{M}{2
\bar{\alpha}a}\right)}{\sqrt{\bar{\alpha}^2
a^2-\frac{M}{\bar{\alpha}a}}}   \,.
\end{equation}
The right hand side term is always positive, therefore we need to
impose the condition $\zeta>-1$ to have a solution. The tangential
surface pressure is given by
\begin{equation}\label{CasimirP}
{\cal P}=\frac{1}{8\pi a}\,\frac{1}{\sqrt{\bar{\alpha}^2
a^2-\frac{M}{\bar{\alpha}a}}}
\left[\left(\frac{2-\zeta}{1+\zeta}\right)\bar{\alpha}^2
a^2-\left(\frac{1-2\zeta}{1+\zeta}\right)\frac{M}{2\bar{\alpha}a}
\right]     \,.
\end{equation}
Equation (\ref{CasimirP}) may also be obtained by substituting
$\sigma=2{\cal P}$ into Eq. (\ref{Pfunctionsigma}), and can be
written in a compact form as ${\cal P}=\frac{\bar{\alpha}}{8\pi}
\, \frac{\Gamma(\xi,\zeta)}{\sqrt{1-\xi^3}}$, with
$\Gamma(\xi,\zeta)$ given by
\begin{equation}\label{GammaCasimir}
\Gamma(\xi,\zeta)=\left(\frac{2-\zeta}{1+\zeta}\right)-
\left(\frac{1-2\zeta}{1+\zeta}\right)\frac{\xi^3}{2} \,.
\end{equation}
Qualitatively, from Fig. 6, we verify that for low values of
$\xi$ (high $a$) and high values of $\zeta$, $\Gamma(\xi,\zeta)$
is negative, implying a surface tension, ${\cal P}<0$. While, for
high values of $\xi$ (low $a$, in the proximity of $r_b$) and
for all $\zeta$, $\Gamma(\xi,\zeta)$ is positive, implying a
surface pressure, ${\cal P}>0$. In addition, one also verifies
that for low values of $\zeta$ and for all $\xi$,
$\Gamma(\xi,\zeta)$ is also positive, implying a surface pressure.
We have ${\cal P}=0$, i.e., a surface boundary (as $\sigma=0$ from
the equation of state $\sigma=2{\cal P}$), for
$\zeta_0=(2-\xi^3/2)/(1-\xi^3)$; a tangential surface pressure,
${\cal P}>0$, for $-1<\zeta<\zeta_0$ and a tangential surface
tension, ${\cal P}<0$, for $\zeta>\zeta_0$.

In relation to the energy conditions, from the equation of
state, $\sigma=2{\cal P}$, one readily verifies that $\sigma \geq
0$, if ${\cal P}\geq 0$. Thus, the WEC is satisfied only if ${\cal
P}\geq 0$. This restriction is verified in the domain $-1<\zeta
\leq \zeta_0$, which is consistent with Eq.
(\ref{energyconddomain}).

\begin{figure}[h]
  \centering
  \includegraphics[width=3.0in]{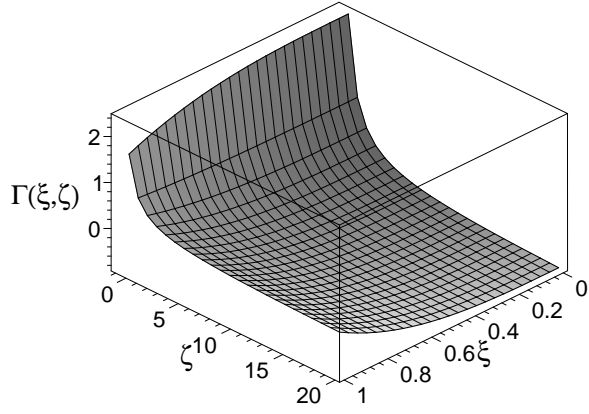}
  \caption{The graph represents the sign of ${\cal P}$ for the
  traceless stress-energy tensor,
  with the equation of state $\sigma -2{\cal P}=0$.
  We have defined $\xi=M^{1/3}/(\bar{\alpha}a)$ and $\zeta=1+a\Phi'(a)$.
  For low values of $\xi$ (high $a$) and high values of $\zeta$,
  $\Gamma(\xi,\zeta)$ is negative, implying a surface tension, ${\cal P}<0$.
  While, for high values of $\xi$ (low $a$, in
  the proximity of $r_b$) and for all $\zeta$,
  $\Gamma(\xi,\zeta)$ is positive, implying a surface pressure, ${\cal P}>0$.
  In addition, one also verifies that for low values of $\zeta$
  and for all $\xi$, $\Gamma(\xi,\zeta)$ is also positive,
  implying a surface pressure.
  See text for details.}
\end{figure}

\subsection{Matching of the equations II: The radial pressure}

To construct specific solutions of planar wormholes with a generic
cosmological constant, the behavior of the radial pressure across
the junction surface $\Sigma$ is required. The pressure balance
equation which relates the radial pressure across the boundary in
terms of the surface stresses of the thin shell \cite{Visser},
deduced in the Appendix, is given by
\begin{eqnarray}\label{behaviorradialpressure}
\left(\bar{p}_r(a)+\frac{3\bar{\alpha}^2}{8\pi} \right) -
\left(p_r(a)+\frac{3\alpha^2}{8\pi} \right) &=&
\frac{1}{a}\,\left(\sqrt{\bar{\alpha}^2 a^2-\frac{M}{\bar{\alpha}
a}}+\sqrt{\alpha^2 a^2-\frac{m(a)}{\alpha a}}\right)\,{\cal P}
      \nonumber       \\
& &-\frac{1}{2} \left(\frac{\bar{\alpha}^2
a+\frac{M}{2\bar{\alpha} a^2}} {\sqrt{\bar{\alpha}^2 a^2-
\frac{M}{\bar{\alpha} a}}}+\Phi'(a)\,\sqrt{\alpha^2
a^2-\frac{m(a)}{\alpha a}} \right) \,{\sigma} \,,
\end{eqnarray}
where $\sigma$ and ${\cal P}$ are given by Eqs.
(\ref{surfenergy})-(\ref{surfpressure}), respectively.
Since $\bar{p}_r(\bar{r})=0$ Eq. (\ref{behaviorradialpressure})
is reduced to
\begin{eqnarray}
p_r(a)&=&\frac{3}{8\pi}(\bar{\alpha}^2-\alpha^2)-
\frac{1}{a}\,\left(\sqrt{\bar{\alpha}^2 a^2-\frac{M}{\bar{\alpha}
a}}+\sqrt{\alpha^2 a^2-\frac{m(a)}{\alpha a}}\right)\,{\cal P}
      \nonumber       \\
& &+\frac{1}{2} \left(\frac{\bar{\alpha}^2
a+\frac{M}{2\bar{\alpha} a^2}} {\sqrt{\bar{\alpha}^2 a^2-
\frac{M}{\bar{\alpha} a}}}+\Phi'(a)\,\sqrt{\alpha^2
a^2-\frac{m(a)}{\alpha a}} \right) \,{\sigma} \,,
\end{eqnarray}
giving the interior radial pressure at the boundary as a function
of the metric and matter fields at the junction surface.


\section{Model of a planar traversable wormhole minimizing the
usage of exotic matter}

One can also construct planar traversable wormhole solutions that
minimize the region of the null energy condition violation. As the
violation of the energy conditions is a particularly problematic
issue, depending on one's point of view, it is of particular
interest to minimize the exotic matter considered at hand
\cite{VKD,Kuhfittig,Kuhfittig2}. For instance, one can confine the
exotic matter to an arbitrarily small region around the throat,
and consider that the surface stresses on the thin shell obey the
energy conditions. The approach of Kuhfittig
\cite{Kuhfittig,Kuhfittig2}, which we will follow, is to choose a
special form function $m(r)$ that can be made to minimize the
violation of the NEC. This yields interestingly enough results for
our planar case, as it has yielded for the spherically symmetric
case presented in \cite{Kuhfittig}.

The interior wormhole solution extends from the throat at $r_0$,
to a radius $a$, where it is matched to an exterior vacuum
solution. To minimize the null energy condition violation, we wish
to deduce a form function, $m(r)$, that somehow yields a solution
arbitrarily close to a solution of $\rho(r)+p_r(r)=0$. To simplify
the analysis, consider a constant redshift function, $\Phi'(r)=0$.
Taking into account Eq. (\ref{NEC}) with $\rho(r)+p_r(r)=0$, we
deduce $m(r)=\alpha^3r^3-k\alpha r$, with $k>0$. With this
motivation, we define the following form function
\begin{equation}\label{specificformfunction}
m(r)=\alpha^3r^3-k\alpha r+\alpha
\,\epsilon(r),   \;\;   {\rm for}\;\; r_0\leq r\leq a  \,,\\
\end{equation}
where $\epsilon$ is a small positive adjustment in the interval
$r_0\leq r \leq a$. To satisfy $m(r_0)=\alpha^3 r_0^3$ at the
throat, the condition $\epsilon(r_0)=kr_0$ is imposed. With the
definition of Eq. (\ref{specificformfunction}), and considering
the particular case of $\epsilon'(r_0)=0$, we also verify that the
flaring-out condition, Eq. (\ref{flarecondition}), is met if
$k>0$.

In the interval $r_0 \leq r \leq a$, from Eqs.
(\ref{rhoWHlambda})-(\ref{lateralpWHlambda}) we have the following
stress-energy functions
\begin{eqnarray}
\rho(r)&=&\frac{1}{8\pi}\;\left(-\frac{k-\epsilon'(r)}{r^2}+3\alpha^2
\right)
        \label{specialrho}       \,,\\
p_r(r)&=&\frac{1}{8\pi}\;\left(\frac{k-\epsilon(r)/r}{r^2}-3\alpha^2
\right)
        \label{specialradialpressure}      \,, \\
p(r)&=&\frac{1}{8\pi}\;\left(\frac{\epsilon(r)-r\epsilon'(r)}{2r^3}-3\alpha^2
\right)     \label{speciallateralpressure}        \,.
\end{eqnarray}
The addition of Eqs. (\ref{specialrho}) and
(\ref{specialradialpressure}) entails the following relationship
\begin{equation} \label{specificformNEC}
\rho(r)+p_r(r)=-\frac{1}{8\pi}\,\frac{\epsilon(r)-\epsilon'(r)r}{r^3}
\,.
\end{equation}
One may now consider $\epsilon'(r_0)=0$. An example satisfying
this condition is
\begin{equation}\label{epsilon}
\epsilon(r)=kr_0\left[1-(r-r_0)^2/(a-r_0)^2 \right],
\end{equation}
(see \cite{Kuhfittig} for a similar example). Thus, Eq.
(\ref{specificformNEC}) at the throat, taking into account
$\epsilon(r_0)=kr_0$ with $k>0$, reduces to
\begin{equation}
\rho(r_0)+p_r(r_0)= -\frac{1}{8\pi}\,\frac{k}{r_0^2} \,.
\end{equation}
Note that $k$ can be made arbitrarily close to zero, and
consequently, the region of the null energy condition violation
can be made arbitrarily small.

Now we have to check that the surface stresses are in line with
the idea of minimizing the NEC. Matching Eq.
(\ref{metricwormholelambda}) with the exterior vacuum solution,
Eq. (\ref{planarBH}), provides the surface stresses of Eqs.
(\ref{surfenergy})-(\ref{surfpressure}). From Eq. (\ref{sigma+P}),
with $\zeta=1$ as $\Phi'(r)=0$, we verify that the NEC is always
verified, i.e.,
\begin{equation}
\sigma +{\cal P}=\frac{1}{8\pi a} \left(\sqrt{\alpha^2
a^2-\frac{m(a)}{\alpha a}}  + \frac{\frac{3M}{2\bar{\alpha}
a}}{\sqrt{\bar{\alpha}^2 a^2-\frac{M}{\bar{\alpha} a}}} \,\right)
\,.
\end{equation}
Here, we consider a positive mass parameter, $M>0$, so that we
have $\sigma +{\cal P}>0$, satisfying the NEC. We can go further
and impose the WEC. In this case, we have $\sigma \geq 0$, i.e.,
$(\alpha^2 a^2-m(a)/\alpha a)\geq (\bar{\alpha}^2
a^2-M/\bar{\alpha}a)$. Taking into account the specific form
function of Eq. (\ref{specificformfunction}), one deduces a
restriction for the mass (per unit area) given by $M \geq
\bar{\alpha}^3 a^3-\bar{\alpha} ak+\bar{\alpha} \epsilon(a)$.
Considering the specific choice of Eq. (\ref{epsilon}), we verify
that $\epsilon(a)=0$ at the junction, so that the restriction
imposed on the mass parameter reduces to $M\geq \bar{\alpha}^3
a^3-\bar{\alpha} ak$. Thus, the WEC holds for $M>\bar{\alpha}^3
a^3$.

In particular, considering a zero surface energy density,
$\sigma=0$, the surface stresses on the thin shell, Eqs.
(\ref{nullsurfenergy})-(\ref{nullsurfpressure}), with $\zeta=1$,
reduce to
\begin{eqnarray}
\sigma_0&=&0   \,,  \\
{\cal P}_0&=&\frac{1}{8\pi a} \,\frac{\bar{\alpha}^2
a^2+\,\frac{M}{2\bar{\alpha}a}}{\sqrt{\bar{\alpha}^2
a^2-\frac{M}{\bar{\alpha}a}}}    \,.
\end{eqnarray}
Thus ${\cal P}$ is always positive  ${\cal P}>0$,, i.e., a surface
pressure, satisfying the WEC and NEC. For $\sigma=0$, considering
Eq. (\ref{epsilon}), the mass (per unit area) reduces to
$M=\bar{\alpha}^3 a^3-\bar{\alpha} ak$. Note, however, that as
$k\rightarrow 0$, $M$ can be arbitrarily close to $\bar{\alpha}^3
a^3$ but not equal, as from the equality case we get an unwanted
event horizon.


\section{Conclusion}

We have constructed plane symmetric wormholes (with planar,
cylindrical and toroidal topologies) in an anti-de Sitter
background. We have determined the surface stresses, analyzed the
sign of the tangential surface pressure, displayed an equation
relating the radial pressure across the junction boundary, and
given a model which minimizes the usage of exotic matter.  We have
found that the construction of these wormholes does not involve a
topology change, and thus the wormhole solution can be considered
a domain wall. As such these wormholes do not allow time travel.
We have not considered in this work solutions with zero or
positive cosmological constants as, for plane symmetry, they yield
solutions with negative total masses.



\appendix

\section{Junction conditions}

We shall use the Darmois-Israel formalism to determine the surface
stresses at the junction boundary \cite{Israel, Papahamoui}.
Consider two spacetimes, $M^{+}$ and $M^{-}$, with metrics $g_{\mu
\nu}^{+}(x_{+}^{\lambda})$ and $g_{\mu \nu}^{-}(x_{-}^{\lambda})$,
defined respectively in the coordinate systems $x_{+}^{\lambda}$
and $x_{-}^{\lambda}$. Assume that these spacetimes, $M^{+}$ and
$M^{-}$, have timelike boundaries $\Sigma^{+}$ and $\Sigma^{-}$,
respectively. The induced metrics on the boundaries are
$g_{ij}^{+}(\xi_{+}^{k})$ and $g_{ij}^{-}(\xi_{-}^{k})$,
respectively, where $\xi_{\pm}^k$ are the intrinsic coordinates on
$\Sigma_{\pm}$. The Darmois-Israel formalism consists in pasting
the spacetimes together, demanding that the boundaries are
isometric, having the same coordinates,
$\xi_{+}^k=\xi_{-}^k=\xi^k$. The identification
$\Sigma^{+}=\Sigma^{-}\equiv \Sigma$ provides a single spacetime
given by $M=M^+ \cup M^-$.

The parametric equation for $\Sigma$ is given by $f(x^{\mu}(\xi
^{i}))=0$, and the respective unit $4$-normal to $\Sigma$ is
provided by $n_{\mu}=\pm | \,g^{\alpha \beta}\, (\partial
f/\partial x^{\alpha})\, (\partial f/\partial x^{\beta}) | ^{-1/2}
\;(\partial f/\partial x^{\mu})$, with $n^{\mu}\,n_{\mu}=+1$. The
intrinsic surface stress-energy tensor, $S_{ij}$, is given by the
Lanczos equations \cite{Israel} in the form
$S^{i}_{\;j}=-\frac{1}{8\pi}\,(\kappa ^{i}_{\;\,j}-\delta
^{i}_{\;j}\kappa ^{k}_{\;\,k})$. For notational convenience, the
discontinuity in the second fundamental form or extrinsic
curvatures is given by $\kappa_{ij}=K_{ij}^{+}-K_{ij}^{-}$. The
second fundamental form is defined as
\begin{equation}
K_{ij}^{\pm}=\frac{\partial x^{\alpha}}{\partial \xi ^{i}} \,
\frac{\partial x^{\beta}}{\partial \xi
^{j}}\,\nabla_{\alpha}^{\pm}\,n_{\beta}=-n_{\gamma}
\left(\frac{\partial ^2 x^{\gamma}}{\partial \xi ^{i}\,\partial
\xi ^{j}}+\Gamma ^{\gamma \pm}_{\;\;\alpha \beta}\;\frac{\partial
x^{\alpha}}{\partial \xi ^{i}} \, \frac{\partial
x^{\beta}}{\partial \xi ^{j}} \right)
   \label{defextrinsiccurvature}     \,.
\end{equation}
The superscripts $\pm$ correspond to the exterior and interior
spacetimes, respectively.

Considerable simplifications occur due to plane symmetry, namely
$\kappa ^{i}_{\;j}={\rm diag} \left(\kappa
^{\tau}_{\;\,\tau},\kappa ^{x}_{\;\,x},\kappa
^{x}_{\;\,x}\right)$. Thus, taking into account the Lanczos
equations, the intrinsic surface stress-energy tensor may be
written as $S^{i}_{\;j}={\rm diag}(-\sigma,{\cal P},{\cal P})$,
where the surface energy density, $\sigma$, and the surface
pressure, ${\cal P}$, are given by
\begin{eqnarray}
\sigma &=&-\frac{1}{4\pi}\kappa ^{x}_{\;\,x} \,,   \label{sigma} \\
{\cal P} &=&\frac{1}{8\pi}(\kappa ^{\tau}_{\;\,\tau}+\kappa
^{x}_{\;\,x})    \label{surfacepressure}  \,.
\end{eqnarray}
This simplifies the determination of the surface stress-energy
tensor to that of the calculation of the non-trivial components of
the extrinsic curvature, or the second fundamental form.

In particular, we shall match the interior and exterior metrics,
Eq. (\ref{metricwormholelambda}) and Eq. (\ref{planarBH}),
respectively. The junction surface is situated at $r=\bar{r}=a$.
In order for these line elements to be continuous across the
junction surface, $ds^2|_{\Sigma}= \bar{ds}^2|_{\Sigma}$, we
consider the following coordinate transformations:
$\bar{t}=t\,e^{\Phi(a)}/\sqrt{\bar{\alpha}^2 a^2-M/\alpha a}\;$,
$\;d\bar{r}/dr|_{r=a}=\sqrt{\bar{\alpha}^2 a^2-M/\alpha
a}/\sqrt{\alpha^2 a^2-m(a)/\alpha a}\;$, $\;\bar{\alpha}
\bar{x}=\alpha x$ and $\bar{\alpha} \bar{y}=\alpha y$.

Using Eq. (\ref{defextrinsiccurvature}), the non-trivial
components of the extrinsic curvature are given by
\begin{eqnarray}
K ^{\tau \;+}_{\;\,\tau}&=&\frac{1}{a}\frac{\bar{\alpha}^2
a^2+\frac{\tilde{M}}{2\bar{\alpha} a}} {\sqrt{\bar{\alpha}^2 a^2-
\frac{\tilde{M}}{\bar{\alpha} a}}}  \;,
         \label{Kplustautau}  \\
K ^{\tau \;-}_{\;\,\tau}&=&\Phi'(a)\sqrt{\alpha^2
a^2-\frac{m(a)}{\alpha a}}  \;, \label{Kminustautau}
\end{eqnarray}
and
\begin{eqnarray}
K^{x\;+}_{\;\,x}&=&K^{y\;+}_{\;\,y}=
\frac{1}{a}\sqrt{\bar{\alpha}^2 a^2-\frac{\tilde{M}}{\bar{\alpha}
a}} \;,
    \label{Kplusx}\\
K^{x \;-}_{\;\,x}&=&K^{y\;-}_{\;\,y}= \frac{1}{a}\sqrt{\alpha^2
a^2-\frac{m(a)}{\alpha a}} \label{Kminusx} \;.
\end{eqnarray}
Thus, the Einstein field equations, Eqs.
(\ref{sigma})-(\ref{surfacepressure}), with the extrinsic
curvatures, then provide us with the following surface stresses
\begin{eqnarray}
\sigma&=&-\frac{1}{4\pi a} \left(\sqrt{\bar{\alpha}^2
a^2-\frac{M}{\bar{\alpha} a}}
-\sqrt{\alpha^2 a^2-\frac{m(a)}{\alpha a}}\;\right)  \label{energy}  \,,\\
{\cal P}&=&\frac{1}{8\pi a} \left[\frac{2\bar{\alpha}^2
a^2-\frac{M}{2\bar{\alpha} a}}{\sqrt{\bar{\alpha}^2
a^2-\frac{M}{\bar{\alpha} a}}}-\sqrt{\alpha^2
a^2-\frac{m(a)}{\alpha a}}\left(1+a\Phi'(a)\right) \,\right]
\label{P} \,.
\end{eqnarray}

One may also obtain an equation governing the behavior of the
radial pressure in terms of the surface stresses at the junction
boundary from the following identity
\begin{equation}\label{pressurebalance}
\left[\,T^{\rm
total}_{\hat{\mu}\hat{\nu}}\,n^{\hat{\mu}}n^{\hat{\nu}}
\right]=\frac{1}{2}(K^{i\;+}_{\;\,j} +
K^{i\;-}_{\;\,j})\,S^{j}_{\;\,i}   \,,
\end{equation}
where $T^{\rm
total}_{\hat{\mu}\hat{\nu}}=T_{\hat{\mu}\hat{\nu}}+g_{\hat{\mu}\hat{\nu}}\,3\alpha^2/8\pi$
is the total stress-energy tensor, and the square brackets denotes
the discontinuity across the thin shell, i.e.,
$[X]=X^{+}|_{\Sigma}-X^{-}|_{\Sigma}$. Equation
(\ref{pressurebalance}) can also be deduced from the normal
component of the conservation of the stress-energy tensor
\cite{Visser}. Taking into account the values of the extrinsic
curvatures, and noting that the pressure acting on the shell is by
definition the normal component of the stress-energy tensor,
$p_r=T_{\hat{\mu}\hat{\nu}}\,n^{\hat{\mu}}n^{\hat{\nu}}$, we
finally have the following pressure balance equation
\begin{eqnarray}\label{Appendixpressurebalance}
\left(\bar{p}_r(a)+\frac{3\bar{\alpha}^2}{8\pi} \right) -
\left(p_r(a)+\frac{3\alpha^2}{8\pi} \right) &=&
\frac{1}{a}\,\left(\sqrt{\bar{\alpha}^2 a^2-\frac{M}{\bar{\alpha}
a}}+\sqrt{\alpha^2 a^2-\frac{m(a)}{\alpha a}}\right)\,{\cal P}
      \nonumber       \\
& &-\frac{1}{2} \left(\frac{\bar{\alpha}^2
a+\frac{M}{2\bar{\alpha} a^2}} {\sqrt{\bar{\alpha}^2 a^2-
\frac{M}{\bar{\alpha} a}}}+\Phi'(a)\,\sqrt{\alpha^2
a^2-\frac{m(a)}{\alpha a}} \right) \,{\sigma} \,,
\end{eqnarray}
which relates the difference of the radial pressure across the
shell in terms of a combination of the surface stresses and the
geometrical quantities. $\sigma$ and ${\cal P}$ are given by Eqs.
(\ref{energy})-(\ref{P}), respectively.

One may obtain Eq. (\ref{Appendixpressurebalance}) in an
alternative manner using directly the Einstein equations. The
analysis is simplified considering two general solutions of Eq.
(\ref{metricwormholelambda}), an interior and an exterior
solution, written out in the coordinate systems, $(t,r,x,y)$ and
$(\bar{t},\bar{r},\bar{x},\bar{y})$, respectively, and matched at
$\Sigma$. Consider the radial component of the Einstein equations,
Eq. (\ref{radialpWHlambda}), written out in both coordinate
systems and note that in the exterior spacetime we have the
following relationships: $\bar{m}=M$ and
$\bar{\Phi}'(a)=\left(\bar{\alpha}^2\,a-
\frac{M}{2\bar{\alpha}a^2}\right)\big/
\left(\bar{\alpha}^2\,a^2-\frac{M}{\bar{\alpha}a}\right)$. Taking
into account the latter relations and the matching of the
solutions given by Eq. (\ref{metricwormholelambda}) and Eq.
(\ref{planarBH}), from the radial component of the Einstein
equations, we deduce the Eq. (\ref{Appendixpressurebalance}).

Working in the same coordinate systems, the continuity of the
first fundamental form, $ds^2|_{\Sigma}= \bar{ds}^2|_{\Sigma}$,
implies that the metric components are continuous, i.e., $g_{\mu
\nu}^{+}=g_{\mu \nu}^{-}$. Thus, the equation governing the
behavior of the radial pressure at $\Sigma$, Eq.
(\ref{Appendixpressurebalance}), reduces to the following form,
$\bar{p}_{0_{r}}(a)+3\bar{\alpha}^2/8\pi=p_{0_r}(a)+3\alpha^2/8\pi
+ 2{\cal P}_0 \, e^{\Phi(a)}/a$, where we have defined
$e^{\Phi(a)}=\sqrt{\bar{\alpha}^2 a^2-M/\bar{\alpha}a}$ and ${\cal
P}_0$ is given by Eq. (\ref{nullsurfpressure}).

\begin{acknowledgments}
This work was partially funded by Funda\c c\~ao para a Ci\^encia e
Tecnologia (FCT) -- Portugal through projects CERN/FIS/43797/2001
and SAPIENS 36280/99.  JPSL thanks Observat\'orio Nacional do Rio
de Janeiro for hospitality.
\end{acknowledgments}



\end{document}